\documentclass[12pt]{JHEP3}                 

\def\be{\begin{equation}}
\def\ee{\end{equation}}
\def\bea{\begin{eqnarray}}
\def\eea{\end{eqnarray}}
\def\ba{\begin{array}}
\def\ea{\end{array}}

\def\part{\partial}

\def\hi{\hat \imath}

\title{YM on the dielectric brane: a D0-brane tale}
\author{Luca Martucci\\ E-mail: \email{Luca.Martucci@mi.infn.it}\\ Dipartimento di Fisica dell'Universit\`a di Milano, Via Celoria 16, I-20133 Milano\\ INFN Sezione di Milano,Via Celoria 16, I-20133 Milano}
\author{Pedro J. Silva\\ E-mail: \email{Pedro.Silva@mi.infn.it} \\ Dipartimento di Fisica dell'Universit\`a di Milano, Via Celoria 16, I-20133 Milano\\ INFN Sezione di Milano,Via Celoria 16, I-20133 Milano}
\abstract{In this letter  we present a derivation, from the D0-brane picture, of the background monopole field and in general of the full dynamics of the Yang-Mills theory on the dielectric D2-brane of Myers. To do this we study the large $N$ limit of the fuzzy sphere relevant to the dielectric solution. In contrast to the usual interpretation, where the commutative D2-brane picture arises directly from the large $N$ limit of the D0-brane picture, we find that a residual non-commutativity must be preserved, in order to make the connection by means of the Seiberg-Witten map.}

\preprint{IFUM-741-FT}

\begin{document}
\section{Introduction}
 
Since the appearance of D-branes in string theory, our understanding of gauge theories has been changed dramatically. D-brane physics delivers a geometrical picture to many phenomena in Yang-Mills (YM) theory and also gives the opportunity to include many aspects of non-perturbative physics. In particular there is a strong correlation between geometric configuration of the D-branes in the bulk space-time and the physics of the YM theory on the world-volume of the branes. For example, in the last years many phenomenological models inspired or based on string theory compactifications with D-brane states, have been proposed as potential candidates of our physical word.

One of the most interesting aspects of D-brane physics is the possibility to describe the hole body of M-theory in terms of a single class of partonic variables, like D0-branes in the BFSS conjecture \cite{bfss} or string bits in matrix string theory \cite{dvv,bs}. In this approach, the physical states of M-theory are the result of interplay of a more fundamental set that corresponds to the partonic degrees of freedom. 

Following this line of thoughts, we should be able to describe a set of D-branes as a coarse grain approximation of the underlying D0-branes or the string bits. One interesting example of the above situation is the Myers effect \cite{mye1}, where a group of Dp-branes can be described from the perspective of many Dq-branes ($p>q$). Note that, in particular, this implies that the corresponding fields living on the Dp-branes should be tractable in terms of the partonic variables. 

To be more precise, let us take as an example the original dielectric effect of Myers. Here we have the possibility to describe the dielectric D2-brane in two different ways: first in term of a single D2-brane, with commutative world-volume and a magnetic flux of monopole charge $N$; second in term of a set of $N$ D0-branes, forming a non-commutative geometry. 

In the original calculation of Myers both descriptions were shown to agree in the large $N$ limit, in the sense that certain observables like the energy, the radius and the Ramond-Ramond (RR) couplings coincide in both pictures up to corrections of order $1/N^2$. The calculation on the D0-brane picture goes as follows: consider the action for $N$ D0-branes in a time-independent case, in the presence of a constant RR field strength 4-form $F^{(4)}_{t123}=-2f\epsilon_{123}$:
\bea
S_{D0}=-\mu_0\lambda^2\int{ dt\;{\rm
Tr}\left({1\over4}[\phi^i,\phi^j][\phi_j,\phi_i] - {i\over
3}F^{(4)}_{tijk}\phi^i\phi^j\phi^k\right)}, 
\label{do}
\eea 
where $\phi^i$ are $u(N)$ matrix-valued scalars. They represent the nine directions $i=1,2,..,9$ transverse to the $D$0-brane. $\mu_0$ is the charge of the
$D$0-brane and $\lambda=2\pi l_s^2$ ($l_s$ is the string length). 
 This potential has a classical extremum corresponding to a 
non-commutative configuration known as fuzzy sphere defined by 
\bea
\phi^i=fL^i\;\;,\;\;[L^i,L^j]=\epsilon_{ijk}L^k
\eea   
where now $i,j,k$ takes value on $1,2,3$, $\phi^i$ represent the non-trivial scalar fields on the D0-brane world-volume theory, $L^i$ are the $su(2)$ Lie algebra generators in an irreducible  $N$ dimensional matrix representation. 
In particular the radius of the above configuration is given by
\bea
\hat{R}=2\pi l_s^2\sqrt{Tr(\phi^i\phi_i)/N}=\pi l_s^2f\sqrt{N^2-1}.
\eea
In the dual picture, we take the Born-Infeld and Chern-Simons action for a single D2-brane in a spherical configuration of radius $R$, and same external fixed RR 4-form $F^{(4)}_{t123}$. Then, to include the effect of the $N$ D0-branes, a gauge field configuration of monopole charge $N$ is added. The minimum of the potential as a function of $R$ is
\bea\label{radius}
R=\pi l_s^2fN.
\eea 
Although Myers found agreement in the large $N$ limit in a set of observables (like the radius of the sphere, explicit sowed here\footnote{We have not showed the calculation for the energy and the RR couplings, and the above discussion gives only some of the final results. For a complete treatment go to the original work of Myers.}), there is an important missing ingredient in the above relation between the dual descriptions. Basically, we don't have a description corresponding to the gauge field (living on the commutative D2-brane) from the D0-brane point of view\footnote{In the original calculation of Myers, the presence of the monopole field is justify by CS coupling arguments and also by t-duality, in the sense that the monopole number correspond to number of diluted D0-branes in the D2-brane.}. In other words, {\em we would like to understand how  the notion of a monopole field and generally the notion of the gauge field living on the D2-brane emerge  from the D0-brane picture}.

There seems to be more than one way to approach this question. T-duality certainly gives a hint as we know that $[\phi_i,\phi_j]\approx F_{ij}$, therefore pointing out a strong relation between non-commutativity and a non-trivial gauge field, but this idea is clearly not satisfactory enough. A more direct approach will be to try to use the definition of gauge theories in non-commutative geometry \cite{connes,madore,Carow-Watamura:1996wg} and impose a monopole structure, following for example \cite{Grosse:1995jt,Balachandran:1999hx},
that in the large $N$ limit recovers the configuration of Myers. Nevertheless, this extra structure appears artificial and seems to 
be imposed ad hoc.

In this letter we will rather search for string theory arguments to disclose the non-commutative description of the gauge field. Inspired by the results presented in \cite{Seiberg:1999vs,Seiberg:2000zk,Alekseev:1999bs,Iso:2001mg}, we will follow a matrix theory approach, together with the Seiberg-Witten map to define the background YM field configuration of the D2-brane appearing in Myers' dielectric effect calculation. As a bonus we will obtain also the description of the fluctuations around this classical configuration and therefore the full gauge theory description.

In the next section, we present a detailed discussion of the large $N$ limit of the dielectric brane, then we study the overlap of the two dual descriptions of the dielectric brane, obtaining the desired correspondence for the gauge field. We conclude with a summary and some remarks regarding physical consequences and possible future generalizations. At last, an appendix containing a more complete derivation of the results is included.       

After this work was completed, our attention was called on a previous work \cite{Karczmarek:2001pn}, in which there is partial overlap 
with our results.

\section{Dielectric Branes and Seiberg-Witten Map}

In order to understand how to describe the gauge field degrees of freedom from the D0-brane point of view, we will first pay some attention to the definition of the large $N$ limit. Note that, if we simply take $N$ to be large, we will retain the non-commutativity in the D0-brane solution. The commutators of the scalar fields $\phi^i$ don't go to zero. Therefore, the resulting D0-brane picture will not be quite the same as the D2-brane picture. On one hand we have a commutative geometry while, on the other, we have a non-commutative geometry (although we should remember that the algebra of functions on the non-commutative solution will be very close to the algebra of functions on the commutative solution). To solve this slight problem, we have to realize that the correct physical limit (where both picture are supposed to overlap), is obtained by taking not only $N$ big but also the RR field strength to be small.
One way to see this is by noticing that the D0-brane description is valid as long as the distance between the D0-branes is smaller than the string scale $l_s$. Thinking on the D0-branes as forming area elements, we get
\bea
{4\pi\hat{R}^2\over N}\ll l_s^2\;\;\Longrightarrow  \;\; f\ll{1\over l_s\sqrt{N}}\ .
\eea
There is also a constraint coming from the D2-brane picture, since we need the radius of the commutative sphere to be larger than the string scale,
\bea 
R\gg l_s\;\; \Longrightarrow  \;\; f\gg{1\over l_sN}\ .
\eea
Using the following ansatz for the RR field strength
\be
f^2=\frac{\beta}{\lambda N^{\alpha}}\ ,
\label{rr}
\ee 
we get for large $N$, that $\alpha$ takes values on the interval $[1,2]$ and $\beta$ satisfies the inequality $N^{\alpha-2}\ll\beta\ll N^{\alpha-1}$. 
The equation (\ref{radius}) for the  radius of the sphere becomes
\bea\label{radius2}
R=\frac{\sqrt{\lambda\beta}}{2}N^{1-\frac{\alpha}{2}}\ .
\eea
The extremal cases are of particular interest as they correspond, in the first case $(\alpha=1,\beta\ll 1)$, to a fuzzy sphere of large radius depending on $N$ with the biggest RR field, in the second $(\alpha=2,N\gg\beta\gg 1)$, to a fuzzy sphere of constant radius. 
 
Once we have clarified the necessary conditions to take the correct limit, we return to the object of this letter, i.e. the description of the gauge field. 

The idea of having a large number of D0-branes forming Dp-brane configurations is not new and can be found in the context of M(atrix) theory or in string theory calculations on brane dynamics. In particular, Seiberg and Witten \cite{Seiberg:1999vs} and Seiberg \cite{Seiberg:2000zk} 
 discuss the background independence of these type of formulations and how to interpolate from the matrix model description of a Dp-brane to the usual commutative Born-Infeld description. They introduce the Seiberg-Witten map defined between non-commutative and commutative D-brane effective theories. In particular, we have the following well known relations,
\bea
G=-\lambda^2 {\cal F}\frac{1}{g}{\cal F}\;\;&,&G_s=g_s det(\lambda {\cal F}g^{-1})^{\frac12} \nonumber \\
\Theta=-\frac{1}{{\cal F}}\;\;&,&\;\;\Phi=-{\cal F}\ ,
\label{sw}
\eea
between open string metric $G$, open string coupling constant $G_s$, noncommutative parameter $\Theta$, two-form $\Phi$ and close string metric $g$, close string coupling constant $g_s$ and constant magnetic field ${\cal F}$. This map is constructed for the flat case and is understood as a field redefinition related to different regularizations of string theory\footnote{Note the different in the sign conventions from the original work. Here we follow the same conventions of Myers i.e. $B \rightarrow -B, A\rightarrow -A, {\cal F}\rightarrow -{\cal F}, \Phi\rightarrow -\Phi$.}.

Now, if the Seiberg-Witten map was defined for curved spaces in RR backgrounds, we could use it to answer our question. Unfortunately this generalization has not yet appeared (to the best of our knowledge). Nevertheless, we are interested in the large $N$ limit, since it is only in this regime that the D0-brane picture and the D2-brane picture should agree, and  if in equations
(\ref{rr}) and (\ref{radius2}) we take $\alpha<2$,  
in this limit the curvature of the sphere and the RR field are very small, hence the use of Seiberg-Witten map seems possible at least at a first order approximation.

Therefore, our plan is to take the correct large $N$ limit and, by means of the Seiberg-Witten map, recover the monopole field on the commutative sphere.

We start with the D2-brane picture and, for simplicity, we focus on a small neighborhood of the  north pole of  the sphere. In spherical coordinates the monopole field takes the form ${\cal F}_{\psi\varphi}=\frac{N}{2}sin\psi$. We introduce the coordinates $\sigma^a,\ a=1,2$, such that,
\begin{equation}
\sigma^1=R\psi cos\varphi\ , \ \sigma^2=R\psi sin \varphi \ . 
\end{equation}
In these coordinates the induced closed string metric in the north pole is the euclidean 
$g_{ab\mid{(\sigma^c=0)}}=\delta_{ab}$, hence in the neighborhood  we take $g_{ab}\simeq\delta_{ab}$. If we now define $\rho =R\psi=\sqrt{(\sigma^1)^2+(\sigma^2)^2}$, the field strength in these coordinates becomes
\bea
{\cal F}_{ab}=\frac{1}{\lambda f \rho} sin(\frac{\rho}{R})\epsilon_{ab}\simeq \frac{2}{\lambda\beta}\epsilon_{ab}\ .
\label{magnetic}
\eea
where we have used equation (\ref{rr}) with  $\alpha=1$ (this is the case were the RR field decreases more slowly)\footnote{In the following, we will restrict to the $\alpha=1$ case such that the quantity 
$fR$, that appears in many expressions, remains finite in the $N\rightarrow \infty$ limit.
The general case needs an extension of the Seiberg-Witten map to curved backgrounds.}. We can see that locally the monopole field looks like a constant magnetic field and the metric looks flat. Therefore we characterize the above neighborhood on the D2-brane by,
\bea
\Big({\cal F}_{ab}=\frac{1}{\lambda\beta}\epsilon_{ab} \;\;,\;\;g_{ab}=\delta_{ab}\;\;,\;\; g_s\Big)\ .
\label{D2-Np}
\eea

Consider next the equivalent calculation in the D0-brane picture. The fuzzy sphere configuration is defined by the matrix-value scalar fields $\phi^i$. The variable with length dimension is recovered by defining $x^i=\lambda\phi^i$ which satisfies the relations
\begin{eqnarray}
[x^i,x^j]=i\lambda f\epsilon_{ijk}x^k\ .
\end{eqnarray}
In the large $N$ limit, to be ``near the north pole'' corresponds to use the equation $x^3\simeq\hat{R}$ \footnote{This statement can be understood from the point of view of coherent states. Coherent states are use to identify quantum functions on the fuzzy sphere with classical functions on the sphere. Here, we are restricting the Hilbert space of the fuzzy sphere to a subspace, such that in the large $N$ limit the coherent states that can be form from this subspace, map quantum operators to classical functions on the north pole. On this subspace $x^3\simeq \hat{R}$.}. Then, the commutation relation adopts the form
\begin{eqnarray}
[x^a,x^b]\simeq i\lambda f \hat{R}\epsilon_{ab}\simeq i{\lambda\beta\over 2}\epsilon_{ab}.\label{angle}
\label{ncf}
\end{eqnarray}
where $a,b=1,2$. In the above limit, we can redefine the fuzzy sphere as a non-commutative geometry with constant theta angle $\Theta_{ab}={\lambda\beta\over 2}\epsilon_{ab}$ in a ``flat manifold'', at least at first order in a $1/N$ expansion. Therefore, we can use the set of equations (\ref{sw}) to rewrite the above non-commutative theory  as a commutative theory with constant magnetic field.

As can be seen from the expressions (\ref{angle}) and (\ref{magnetic}) and the Seiberg-Witten map, we get a perfect agreement in the large $N$ limit between the monopole field in the D2-brane and the  non-commutativity of the D0-brane. Our first conclusion is that {\em the non-commutative space itself is translated into the background gauge field}. 

At this point we have been able to recover the constant magnetic monopole field characteristic of the D2-brane from the D0-brane point of view, to go one step further in this derivation, we consider general fluctuations around the dielectric solution,
\bea
\lambda\phi_i= x_i+\lambda f\hat{A}_i\;=\; \lambda f(L_i +\hat{A}_i) 
\label{exp}
\eea
where $x_i$ corresponds to the fuzzy solution of Myers, $\hat{A}_i$ are the fluctuations and as before $L_i$ correspond to the Lie algebra generators of $su(2)$. In the usual commutative case, if we consider a non-Abelian gauge field on the sphere we get the following form for the field strength, $F=dA+iA\wedge A$. Using the bases $\{l^i\}_{i=1}^3$ of 1-forms dual to the vector fields which generate the rotations $l_i$ ($[l_i,l_j]=-\epsilon_{ijk}l_k$), we obtain, 
\begin{equation}
F_{ij}=l_i(A_j)-l_j(A_i) +i[A_i,A_j] +\epsilon_{ijk}A_k\ .
\end{equation}
To generalize this definition to the non-commutative case one has simply to make the substitution $l_i\rightarrow i[L_i,\cdot]$, obtaining\footnote{Note that we started from Cartesian coordinates, but we have ended in the base $(l_1,l_2,l_3)$, that corresponds to an angular-type base and so it seems that the ``fuzzyfication'' produce a transmutation of the meaning of the indexes.}
\begin{equation}
\hat{F}_{ij}=i[L_i,\hat{A}_j]-i[L_j,\hat{A}_i] +i[\hat{A}_i,\hat{A}_j] +\epsilon_{ijk}\hat{A}_k\ .
\end{equation}
This expressions can be also obtained by applying the more rigorous formulation a la Connes on the fuzzy sphere \cite{Carow-Watamura:1996wg}.
To obtain the effective action of the perturbation $\hat{A}_i$, we substitute the expansion (\ref{exp}) into the D0-brane action (\ref{do}) using the above definitions. In particular, from the first term of the action we get,
\bea
&&Tr \lambda^4\delta^{ij}\delta^{hk}[\phi_i,\phi_h][\phi_j,\phi_k]=\cr
&&=-\lambda^4 Tr f^4 \delta^{ij}\delta^{hk}(\hat{F}_{ih}-\epsilon_{ihm}L_m)(\hat{F}_{jk}-\epsilon_{jkr}L_r)+ more...
\eea
Then we can replace the trace $Tr$ in the D0-brane action by the symbol $(N/4\pi)\int d\Omega$ (that in the large $N$ limit corresponds the the notion of integration on the sphere). 
Following the analysis of Seiberg presented in \cite{Seiberg:2000zk},
we can extract\footnote{See the appendix for the complete derivation} the form of the effective metric $G_{ij}$, the 2-form $\Phi_{ij}$ and the string coupling constant, giving
\bea
\Big(G_{ij}=\frac1{f^2}\delta_{ij}\;\;,\;\;\Phi_{ij}=-\epsilon_{ijk}L^k\;\;,\;\; G_s={g_s\over f\hat{R}} \Big) .
\eea
Next, we take the large $N$ limit and focus on the ``north pole'' sector of the theory, where we find that $L_3=x_3/\lambda f \simeq \hat{R}/\lambda f$, and we change to the ``almost flat'' non-commutative coordinates $x^a$, $a=1,2$ introduced in (\ref{ncf}), obtaining
\bea
&&\Phi_{ab}=-{1\over \hat{R}^2}\epsilon_{ab}L_3\simeq -{2\over \lambda\beta}\epsilon_{ab},\nonumber \\
&&G_{ab}={1\over \hat{R}^2f^2}\delta_{ab}\simeq {4\over \beta^2}\delta_{ab},\nonumber \\
&&G_s={g_s\over f\hat{R}}\simeq {2g_s\over \beta}
\eea  
then, using the Seiberg-Witten map in the above non-commutative gauge theory, we get a commutative gauge theory with constant magnetic field in full agreement with the expressions of equation (\ref{D2-Np}). In the appendix we give a more complete derivation, using the full time dependent BI action, where the YM dynamics of the D2-brane is recovered. 

To finish this discussion, we mention that the above action can be rewritten in terms of $\hat{A}_i$ and $L_i$ only, giving
\bea
-T_0\lambda^2f^4Tr\left\{{1\over 4}\hat{F}_{ij}\hat{F}^{ij}-{1\over2}\epsilon^{ijk}(i[L_i,\hat{A}_j]\hat{A}_k+{i\over3}\hat{A}_i[\hat{A}_j,\hat{A}_k]+{1\over2}\epsilon_{ijl}\hat{A}^l\hat{A}_k)\right\}
\eea
which is the well known action corresponding to non-commutative YM theory plus a Chern-Simon term in the fuzzy sphere 
\cite{Klimcik:1999uk}. It arises naturally also in a CFT approach to brane dynamics
in background fluxes \cite{Alekseev:1999bs} and is used in matrix models \cite{Iso:2001mg}. Therefore, we have not only recovered the monopole field of the D2-brane, but also the dynamics of the YM theory, since we have shown that the non-commutative YM theory on the fuzzy sphere, in the large $N$ limit near the ``north pole'', goes to a non-commutative YM theory with a 2-form $\Phi$ that after the use of the Seiberg-Witten map gives the commutative YM theory of the D2-brane. 

Actually, we should say that string theory is also telling us how to define YM theory on the fuzzy sphere. Although there are a few different constructions of YM in the fuzzy sphere, all these programs face the problem of the appearance of an extra degree of freedom in the gauge field related to a ``radial direction''. In the above construction, we still have that same radial mode but this time we interpreted this ``scalar field'' (in the large $N$ limit) as corresponding to the radial fluctuation of the D2-brane. For example we can take the gauge field defined in (\ref{exp}), and consider a splitting of the form
\bea
\hat{A}_i=\hat{a}_i+n_i\phi_r,
\eea
where $n_i=x_i/|x|$ and $n_i\hat{a}^i=0$. Note that we define the radial mode from the action on the left only i.e. $n^i\hat{A}_i=\phi_r$ (there are more symmetric prescriptions, but in the large $N$ limit we anyway recover the same commutative restriction $n^ia_i =0$ characteristic of a gauge field on the sphere). In the large $N$ limit, the field $a_i$ corresponds to the usual gauge field, while the scalar field $\phi_r$ becomes the radial fluctuation of the D2-brane. In fact, we can check that the infinitesimal $U(N)$ transformation of the original D0-brane matrix model $\delta_{\rho}\phi_i=i[\phi_i,\rho]$ corresponds for the fields $(\hat{a}_i,\phi_r)$ to 
\bea
&&\delta_{\rho} \hat{a}_i=i(\delta_i^j-n_in^j)[L_j+\hat a_j,\rho]+i(\delta_i^j-n_in^j)[n_j\phi_r,\rho] \nonumber \\
&&\delta_{\rho} \phi_r=n^i(i[L_i,\rho]+i[\hat a_i\rho]+ i[n_i\phi_r,\rho]).
\eea 
where both transformations go to the usual non-commutative gauge and scalar transformation in the large $N$ limit on the ``north pole''. Therefore we have recovered the missing ingredients to complete the world-volume theory of the D2-brane \footnote{Note that only in the large $N$ limit the non-commutative field and the radial fluctuation can be separated; in the finite $N$ case,
it makes no sense to use the splitting.}.

\section{Conclusions and Discussions}

In this short letter we have shown explicitly how the full YM dynamics on the dielectric D2-brane can be described from the D0-brane point of view. To do this, we have studied in detail the large $N$ limit in the construction of Myers, founding for the consistency of this solution a scaling dependent on $N$ in the external RR field. Because of this scaling, starting from the fuzzy sphere solution, we were able to end up (in the large $N$ limit) in a non-commutative geometry, reminiscence of the original fuzzy sphere, where the Seiberg-Witten map can be applied to obtain a commutative theory. This final theory matches the field content of the corresponding D2-brane. As a bonus, we have seen how string theory solves the problem of the ``radial mode'' present in the definition of non-commutative YM theory in the fuzzy sphere, in the sense that this mode should be understood as the radial fluctuation of the D2-brane.

We have also included an appendix where the matching between the two dual pictures for the gauge field is shown in details including time-dependent perturbations and for the full action, not only the leading order expansion used in (\ref{do}). Although Myers used a similar analysis with the full potentials for the breathing radial mode, it is still intriguing that we can find the correspondence at all orders in our complementary case. It remains to give a description of the duality for the full radial perturbation.

Moreover, in this letter we have approximated the non-commutative space ``fuzzy sphere'' with a non-commutative plane, in the limit of large $N$. Hence, we were able to use the Seiberg-Witten map that is only defined for flat space-time or toroidal compactifications. It will be very interesting to generalize this map to  quantum compact spaces, hence the relation between the D0-brane picture and the D2-brane picture could be obtained globally and for the full dynamics of the YM and radial fluctuations. Nevertheless, it seems that the generalized map will produce a commutative theory with ultraviolet and infrared cut-off since the quantum theory will have finite degrees of freedom\footnote{There is a generalization of the Seiberg-Witten map for the fuzzy sphere, but uses a modify start-product \cite{Hayasaka:2002db}.}. A deeper analysis goes beyond this work so we postpone it for future work.

Finally, we should say that although the above analysis was carried on for the dielectric effect, there are other very similar situations in M-theory where the results will carry on, like in M(atrix) theory or matrix string theory (where this time the D-branes are made 
of string bits \cite{Silva:2001ja}) or even in the case of D-branes on compact group manifolds, where the non-commutative geometry appears from the CFT analysis and the corresponding relation with the D0-brane picture is also explicit \cite{Alekseev:1999bs,Schomerus:2002dc}.

\acknowledgments

We thank G. Alexanian and D. Zanon for useful discussions. This work
was partially supported by INFN, MURST and by the European Commission RTN program HPRN-CT-2000-00131, in association with the University of Torino. 

\appendix
\section{A More Complete Calculation}

In this appendix we generalize the discussion of the previous section both working with a time dependent perturbation of the fuzzy solution and using  the full D0-brane action, in contrast with the calculation showed in the main body of the article. We start our analysis from the non-Abelian action of Myers.

The first part of the non-Abelian D0 effective action is the Born-Infeld term
\bea
S_{BI}=-T_0 \int dt \, STr\left( e^{-\phi}
\sqrt{-\left( P\left[E_{ab}+E_{ai}(Q^{-1}-\delta)^{ij}E_{jb}
\right]\right) \, det(Q^i{}_j)} \right)
\label{eq:1}
\eea
with
\be
 E_{AB} = G_{AB}+ B_{AB}\  \qquad { \rm and}\qquad
Q^i{}_j\equiv\delta^i{}_j + i\lambda\,[\phi^i,\phi^k]\,E_{kj}\ .
\label{eq:2}
\ee
In writing (\ref{eq:1})
we have used a number of conventions taken from Myers \cite{mye1}:
\begin{itemize}

\item{}
Indexes to be pulled-back to the world-line (see below)
have been labeled by $a$.  For other indexes,
the symbol $A$ takes values in the full set of space-time coordinates while
$i$ labels only directions perpendicular to the center of mass world-line.

\item{}
The center of mass degrees of freedom decouple completely and are
not relevant to our discussion. The fields $\phi^i$ thus take
values in the adjoint representation of SU(N). As a result, the
fields satisfy $Tr \phi^i=0$ and form a non-Abelian generalization
of the coordinates specifying the displacement of the branes from
the center of mass. These coordinates have been normalized to have
dimensions of $(length)^{-1}$ through multiplication by
$\lambda^{-1}$.
\end{itemize}

The rest of the action is given by the
non-Abelian Chern-Simon terms. These involve the non-Abelian
`pullback' $P$ of various covariant tensors to the world-volume of
the D0-brane e.g. $P[C_t^{(1)}] = C^{(1)}_\mu D_t x^\mu=
 C^{(1)}_0 + \lambda  C^{(1)}_i D_t
\phi^i$, where we have used the static gauge $x^0=t,
x^i = \lambda \phi^i$ for a coordinate $x$ with origin at the D0-brane
center of mass.
The symbol $STr$ will be used to denote a trace
over the $SU(N)$ index with a complete symmetrization over the
non-Abelian objects in each term.  In this way, the Chern-Simons
terms may be compactly written
\bea S_{CS}=\mu_0\int dt
STr\left(P\left[e^{i\lambda\,\hi_\phi \hi_\phi} ( \sum
C^{(n)}\,e^B)\right]\right)\ .
\eea

The symbol $\hi_\phi$ is a non-Abelian generalization of the
interior product with the coordinates $\phi^i$,
\bea
\hi_\phi \left(\frac{1}{2}C_{AB}dX^AdX^A\right) =
\Phi^iC_{iB}dX^B.
\eea

In our case we have $N$ $D0$ branes (the generalization to Dp branes which polarize into $D(p+2)$ branes is straightforward) in flat space with a constant RR field strength $F^{(4)}_{t123}=-2f\epsilon_{123}$.
Focusing on the first four directions of space-time for the BI part of the action, we have
\begin{equation}
S_{BI}=-T_0\int dt STr\sqrt{-det[ -1 + \lambda^2 D_t \phi_i (Q^{-1})_{ij} D_t \phi_j ]det(Q_{ij})}\ ,
\label{bif2}
\end{equation}
where  $Q_{ij}=\delta_{ij}+i\lambda[\phi_i,\phi_j]$, $D_t\phi_i=\partial_t\phi_i +i[\hat A_t,\phi_i]$ and $i,j$ takes values in $\{1,2,3\}$. Observing that the $U(N)$ connection with zero spatial dimension $\hat A _t(t)$ can be seen as a $U(1)$ connection on the fuzzy sphere, by substituting $\phi=f(L_i+\hat A _i)$, we get
\begin{equation}
D_t\phi_i=f\{ \partial_t -i[L_i,\hat A _t]+i[\hat A_t,\hat A_i]\}=f\hat F_{ti}\ .
\end{equation}
After some manipulations, this action can be rewritten as 
\begin{equation}
S_{BI}=-T_0 f^2\int dt STr\sqrt{-f^2det\{ -G_{\mu\nu} + \lambda[ \hat F _{\mu\nu}-\delta_\mu^i\delta_\nu^j
\epsilon_{ijk}(L_k+A_k)]\} }\ ,
\end{equation}
where $\mu,\nu=0,...,3$ and we have defined the effective metric $G_{\mu\nu}$ 
\begin{equation}
G_{\mu\nu}=\left( \begin{array}{cc}
1 & 0  \\
0 & \frac{\delta_{ij}}{f^2} \end{array}\right)\ .
\end{equation}
Making the formal substitution $STr\rightarrow \frac{N}{4\pi}\int d\Omega_2$, which is valid only in the $N\rightarrow \infty$ limit (here $d\Omega_2$ is to be understood as the volume form on the sphere of radius one), we finally write
\begin{equation}
 S_{BI}=-T_2(G_s)\int dt\int d\Omega_2  \sqrt{-f^2det\{ -G_{\mu\nu} + \lambda[ \hat F _{\mu\nu}-\delta_\mu^i\delta_\nu^j
\epsilon_{ijk}(L_k+A_k)]\} }\ ,
\label{bif}
\end{equation}
where we have borrow the definition $R={\lambda fN\over 2}$ from the D2-brane physics to define
\begin{equation}
T_2(G_s)=\frac{2\pi}{(2\pi l_s)^3 G_s}\ ,\ G_s=\frac{g_s}{fR}\ .
\end{equation}
Note that, there is a disturbing factor of $f^2$ in front of the determinant $det$. Nevertheless, this factor is due to the fact that the fuzzy sphere construction force us to work with a three dimensional embedding space. Recalled that we expect to have a redefinition of the metric only along the tangential directions of the sphere, but in the metric $G_{\mu\nu}$ we have multiplied all three spatial direction by $f^{-2}$.   

To make contact with the D2-brane picture, we take the north pole approximation. Note that in this way we can identify the transverse fluctuation of the fuzzy sphere with the third component of the gauge field (more precisely $\varphi\simeq f\hat A_3$). For simplicity we will set this transverse fluctuation to zero. Using the fact that in this approximation $L_3\simeq R/\lambda f$, we introduce the ``almost flat'' non-commutative coordinates $\sigma^A=(t,\sigma^a)$, $A=0,1,2,\ a=1,2$ that satisfy
\begin{equation}
[\sigma^A,\sigma^B]=i\Theta^{AB}\ , \Theta^{AB}=\frac{\lambda\beta}{2}\delta_A^a\delta_B^b\epsilon_{ab},
\end{equation}
Therefore, the action (\ref{bif}) can now be written in in the following way
\begin{equation}
 S_{BI}=-T_2(G_s)\int d^3\sigma  \sqrt{-det\{ G_{AB} + \lambda[ \hat F _{AB}+\Phi_{AB}]\} }\ ,
\end{equation}
where the effective metric is 
\begin{equation}
G_{AB}=\left( \begin{array}{cc}
-1 & 0 \\
0 & \frac{\delta_{ab}}{R^2 f^2} \end{array} \right)\ ,
\end{equation}
we have introduced the background field
\begin{equation}
{\bf \Phi}_{AB}=-\frac{1}{\lambda f R}\delta_A^a\delta_B^b\epsilon_{ab}\ ,
\end{equation}
and we have used the star product associated to $\theta^{AB}$ in the definition of the action.
 
In these form the BI action is identical to the non-commutative BI action of Seiberg and Witten \cite{Seiberg:1999vs} and so it is equivalent to the commutative BI action
\begin{equation}
 S=-T_2\int d^3\sigma  \sqrt{-det\{ g_{AB} + \lambda[ F _{AB}+{\cal F}_{AB}]\} }\ ,
\end{equation}
where ${\cal F}_{AB}=\frac{1}{\lambda f R}\delta_A^a\delta_B^b\epsilon_{ab}$ is the right background monopole field of the D2-brane as seen in the north pole approximation and $g_{AB}=diag(-1,1,1)$ is the flat metric.  

To complete our analysis we must consider also the CS term of the D0-branes, that can be rewritten as \begin{eqnarray}
S_{CS}&=&-\frac{\mu_0\lambda^2 f^4}{3}\int dt Tr L_iL_i +\mu_0\lambda^2 f^4\int dt Tr (L_i+A_i)(L_i+A_i)-\cr
&&-\mu_0\lambda^2 f^4\int dt Tr \epsilon^{ijk}\{i\hat A_i [L_j,\hat A_k]+
\frac{i}{3}\hat A_i [\hat A_j,\hat A_k]+\frac12 \epsilon_{jkm}\hat A_i\hat A_m\}\ .
\end{eqnarray}
In the north pole approximation with zero transversal fluctuation, the only term that survives is the vacuum energy term
\begin{equation}
S_{CS}\simeq \frac{2\mu_0 \lambda^2 f^4}{3} Tr(L_iL_i)\int dt=\frac{2\mu_0 f^2 \hat{R}^2 N}{3}\int dt\ ,
\end{equation}
which indeed corresponds to the CS term coming from the dual $D2-$brane description (up to correction of order  $O(1/N)$, with no radial fluctuations).

summarizing we have found complete agreement in the north pole approximation between the D0-brane picture and the D2-brane picture for the full action. 

We conclude observing that, because $f\sim 1/\sqrt{N}$ and $\phi_i=f(L_i+\hat{A}_i)$, in the large $N$ limit we can consider only the first terms in the expansion of the D0-brane BI action (\ref{bif2}) such that
\begin{eqnarray}
S &=& S_{BI}+S_{CS}=\cr
 &=& -T_2(G_s)\int dt d\Omega_2 \sqrt{-f^2det{G_{\mu\nu}}} -\cr
&&-\frac{ T_2(G_s)\lambda^2}{4}\int dt 
d\Omega_2 \sqrt{-f^2det{G_{\mu\nu}}} G^{\mu\nu}G^{\rho\sigma}\hat F_{\mu\rho} \hat F_{\nu\sigma}+\cr
&&+\frac{\mu_2(G_s)\lambda^2}{2}\int dt d\Omega_2 \sqrt{-f^2det{G_{\mu\nu}}} G^{li}G^{mj}G^{nk}\epsilon_{lmn}\{i\hat A_i [L_j,\hat A_k]+\cr
&&+\frac{i}{3}\hat A_i [\hat A_j,\hat A_k]+\frac12 \epsilon_{jkm}\hat A_i\hat A_m\}\ ,
\end{eqnarray}
where $\mu_2(G_s)=T_2(G_s)$. 

This action is a generalization of the non-commutative YM theory in the fuzzy sphere for time dependent gauge field.



\end{document}